\documentclass[12pt]{iopart}
\usepackage{epsfig}
\usepackage{iopams}
\newcommand{\eq}{\begin{eqnarray}}
\newcommand{\en}{\end{eqnarray}}

\def\cpv{C\hspace{-0.2em}P\hspace{-0.6em}/ }

\begin{document}

\title{The neutron electric dipole form factor
in the perturbative chiral quark model}
\author{Claudio \ Dib\dag,
Amand Faessler\ddag,
Thomas Gutsche\ddag, \\
Sergey Kovalenko\dag,
Jan Kuckei\ddag,
Valery E. Lyubovitskij\ddag,  \\
Kem Pumsa-ard\ddag}

\address{
\dag \
Departamento de F\'\i sica, Universidad
T\'ecnica Federico Santa Mar\'\i a, \\
Casilla 110-V, Valpara\'\i so, Chile}
\address{
\ddag \
Institut f\"ur Theoretische Physik, Universit\"at T\"ubingen, \\ 
Auf der Morgenstelle 14, D-72076 T\"ubingen, Germany}

\ead{cdib@fis.utfsm.cl,
amand.faessler@uni-tuebingen.de,
thomas.gutsche@uni-tuebingen.de,
Sergey.Kovalenko@usm.cl,
kuckei@tphys.physik.uni-tuebingen.de,
valeri.lyubovitskij@uni-tuebingen.de,
pumsa@tphys.physik.uni-tuebingen.de}

\begin{abstract}
We calculate the electric dipole form factor of the
neutron in a per\-tur\-bative chiral quark model,
parameterizing CP-violation of generic origin by means of
effective electric dipole moments of the constituent quarks
and their CP-violating couplings to the chiral fields.
We discuss the relation of these effective parameters to
more fundamental ones such as the intrinsic electric and
chromoelectric dipole moments of quarks and the Weinberg
parameter. From the existing experimental upper limits on
the neutron EDM we derive constraints on these CP-violating
parameters.
\end{abstract}

\vskip 1cm


\vskip .5cm


\newpage

\section{Introduction}

The origin and manifestations of CP violation have been important
issues in particle physics since its discovery in neutral kaon
decays in 1964 \cite{Christenson:1964fg}. So far CP violation has
been observed only in $K$ and $B$ hadron
decays~\cite{Eidelman:2004wy}, results that are consistent with
the Kobayashi-Maskawa formulation of a complex quark mixing matrix
\cite{Kobayashi:1973fv}, within the Standard Model~(SM) of
electroweak interactions. However, CP violation is not only
important in hadron decays but it is also an inevitable ingredient
for the generation of the observed baryon asymmetry of the
Universe \cite{Sakharov:1967dj}. Yet, the observed CP violation
within the SM is by far insufficient to explain it. This evidence
is pointing to physics beyond the SM, whose contribution to CP
violation in the K and B systems may be irrelevant but should be
dominant for the Universe's baryon asymmetry. Thus, there are
powerful motivations to study all possible CP odd observables and
distinguish its underlying sources. Among those observables, a
great deal of effort has been directed to the study of the
electric dipole moment (EDM) of leptons, neutrons and neutral
atoms. Both SM and various non-SM sources of CP violation have
been considered (for recent reviews see e.g.
Ref.~\cite{rev-EDM,Erler:2004cx}). These studies have been
particularly stimulated by the expectation of great improvements
(2 to 4 orders of magnitude) in the experimental sensitivities to
EDMs in the next decade (for review see Ref.~\cite{Erler:2004cx}).

As it is well known, there are two sources of CP-violation within
the SM: the complex phase of the Cabibbo-Kobayashi-Maskawa (CKM)
matrix in the weak interaction sector, and the $\theta$-term in
the strong interaction
sector~\cite{Belavin:1975fg}-\cite{Crewther:1978zz}. On the one
hand, the complex phase of the CKM matrix is well established and
provides a consistent explanation of the observed CP odd effects
in hadron decays, but predicts an imperceptible contribution to
the EDMs, far below the sensitivity of present or foreseeable
experiments. Indeed, the CKM prediction for the neutron EDM,
$d_n$, ranges from $10^{-31}$ to $10^{-33}$ $e
\cdot$cm~\cite{Shabalin:1982sg} while its present experimental
upper limit~\cite{Harris:1999jx} is
\begin{eqnarray}\label{EDM-exp-1}
|d_n| < 0.63 \times 10^{-25} \  e \cdot{\rm cm} \,.
\end{eqnarray}
On the other hand, the presence of a $\theta$-term is still an open
issue; it leads to a sizable neutron electric dipole
moment~\cite{Baluni:1978rf,Crewther:1979pi} and may significantly
contribute to atomic EDMs, while it is insignificant for CP
violation in hadron decays. Indeed, the non-observation of EDM for
the neutron imposes a very strict upper bound on the value of
$\theta$, of the order of $10^{-10}$. This unnaturally small value
of $\theta$, which is otherwise not restricted by theory, is
called the {\it strong CP problem}. One elegant solution was
proposed by Peccei and Quinn (PQ)~\cite{Peccei:1977hh}, which
makes the $\theta$ parameter to vanish dynamically.
Other important contributions to the atomic and neutron EDMs may
arise from physics beyond the SM.

In the calculations of atomic and neutron EDMs one faces the
problem of translation of the CP violation mechanism formulated at
the quark-gluon level, to the processes at the hadronic or atomic
level. This translation must resort to hadronic and nuclear models
and perform a careful treatment of the effects from the cloud of
atomic electrons. The validity of the estimates then depends on
the validity of these models.

The purpose of the present work is to use the PCQM model to
calculate the neutron EDM and electric dipole form factor (EDFF)
induced by the standard CP-odd QCD $\theta$-term as well as by
generic CP-violating effective interactions arising from physics
beyond the SM. The latter correspond to effective quark-photon and
quark-gluon interactions induced by intrinsic electric and
chromoelectric dipole moments of the constituent quarks and the
Weinberg gluon operator at a more fundamental level. In the
present paper we do not discuss concrete mechanisms for the
generation of these CP-odd interactions, but concentrate on the
hadronic structure aspects of the derivation of the neutron EDM in
terms of these generic CP-violating parameters at low energies.

The problem of the neutron EDM has been studied within different
theoretical approaches: current algebra and chiral perturbation
theory~\cite{Crewther:1979pi,Pich:1991fq,Borasoy:2000pq,Hisano:2004tf},
chiral quark
models~\cite{Musakhanov:1984qy}-\cite{McGovern:1992ix}, lattice
QCD~\cite{Aoki:1989rx,Shintani:2005xg}, QCD sum
rules~\cite{Khatsimovsky:1987fr,Pospelov:1999mv}, an approach
based on solutions of Schwinger-Dyson and Bethe-Salpeter
equations~\cite{Hecht:2001ry}, etc. We use the PCQM which
describes baryons as bound states of three relativistic valence
quarks confined in a static potential and supplemented by a cloud
of pseudoscalar Goldstone bosons, as required by chiral symmetry.
This model has already been successfully applied to the charge and
magnetic form factors of baryons, ground state masses of baryons,
the electromagnetic $N$-$\Delta$ transition, meson-nucleon
sigma-terms and other baryon
properties~\cite{Lyubovitskij:2001nm}.

This paper is organized as follows. Sect.~\ref{s2} contains a
brief review of the PCQM. In Sect.~\ref{s3} we apply the PCQM
calculation of the EDFF and the EDM of the neutron induced by the
individual EDMs of the quarks and their CP-violating couplings to
the chiral fields. In Sections~\ref{s4} we extract constraints for
the corresponding quark EDMs and CEDMs from the existing
experimental limits on the neutron EDM and discuss the
phenomenological implications of our results.

\section{The Perturbative Chiral Quark Model}
\label{s2}

The basis of the perturbative chiral quark model
(PCQM)~\cite{Lyubovitskij:2001nm} is an effective chiral Lagrangian
describing the valence quarks of baryons as relativistic fermions
moving in an external field (static potential)
\begin{equation}
V_{\rm eff}(r)=S(r)+\gamma^0 V(r) , \quad {\rm with}\ r=|\vec x| ,
\label{potential}
\end{equation}
which in the SU(3)-flavor version are supplemented by a cloud of
Goldstone bosons $(\pi, K, \eta)$. Treating Goldstone fields as
small fluctuations around the three-quark core, the linearized
effective Lagrangian is written as:
\begin{eqnarray}\label{L_eff}
{\cal L}_{\rm eff}(x) &=& \bar q(x) [i \not\!\partial - S(r) -
\gamma^0 V(r)] q(x) + \frac{1}{2} \sum\limits_{i=1}^8
[\partial_\mu \Phi_i(x)]^2 \nonumber \\ &+& {\cal L}_I^{str
(1)}(x) + {\cal L}_{\chi SB}(x),
\end{eqnarray}
where we explicitly defined the linearized coupling of Goldstone
bosons to quarks as:
\begin{equation}\label{L_eff_1}
{\cal L}_I^{str(1)}(x) =
- \bar q(x) i\gamma^5 \frac{\hat\Phi(x)}{F} S(r) q(x) \,.
\end{equation}
The additional term ${\cal L}_{\chi SB}$ in Eq.~(\ref{L_eff}) contains
the mass contributions both for quarks and mesons, which explicitly
break chiral symmetry:
\begin{equation}
{\cal L}_{\chi SB}(x) = -\bar q (x) {\cal M} q(x)
- \frac{B}{2} Tr [\hat \Phi^2(x)  {\cal M} ] \,.
\end{equation}
Here, $\hat \Phi = \sum\limits_{i=1}^8 \Phi_i \lambda_i$
is the octet matrix of pseudoscalar mesons,
$F=88$ MeV is the pion decay constant in the chiral
limit, ${\cal M}={\rm diag}\{\hat m,\hat m,m_s\}$
is the mass matrix of current quarks (we restrict to the isospin
symmetry limit $m_u=m_d=\hat m$) and $B=-<0|\bar u u|0>/F^2$ is
the quark condensate constant.
We rely on the standard picture of chiral symmetry
breaking and for the masses of pseudoscalar
mesons we use the leading term in their chiral expansion
(i.e. linear in the current quark mass):
\begin{eqnarray}
M_{\pi}^2=2 \hat m B, \hspace*{.5cm} M_{K}^2=(\hat m + m_s) B,
\hspace*{.5cm} M_{\eta}^2= \frac{2}{3} (\hat m + 2m_s) B  \,.
\nonumber
\end{eqnarray}
In our analysis we use the following set
of parameters:
\begin{equation}
\hat m = 7 \;{\rm MeV},\; m_s =25 \hat m,\;
B = M^2_{\pi^+}/ 2 \hat m = 1.4 \;{\rm GeV} \,.
\end{equation}
The meson masses satisfy the Gell-Mann-Oakes-Renner and
Gell-Mann-Okubo relations. In addition, the linearized effective
Lagrangian fulfills PCAC. The properties of baryons, which are
modeled as bound states of valence quarks surrounded by a meson
cloud, are then derived using perturbation theory. At zeroth
order, the unperturbed Lagrangian simply describes a nucleon as
three relativistic valence quarks which are confined by an
effective one-body static potential $V_{\rm eff}(r)$ in the Dirac
equation. We denote the unperturbed three-quark ground-state as
$|\phi_0 \rangle$, with the normalization $\langle \phi_0 | \phi_0
\rangle = 1$. We expand the quark field $q$ in the basis of
eigenstates generated by this potential as
\begin{eqnarray}\label{q-expand}
q(x) = \sum_{\alpha} b_\alpha u_\alpha(\vec{x})
\exp(-i{\cal E}_\alpha t)
\end{eqnarray}
where the quark wave functions $\{ u_\alpha \}$ in orbits $\alpha$
are the solutions of the Dirac equation including the potential
$V_{\rm eff}(r)$. The expansion coefficients $b_\alpha$ are the
corresponding single quark annihilation operators. All
calculations are performed at an order of accuracy ${\cal O}
(1/F^2,\hat{m},m_s)$. In the calculation of matrix elements, we
project the quark diagrams on the respective baryon states. The
baryon states are conventionally set up by the product of ${\rm
SU(6)}$ spin-flavor and ${\rm SU(3)_c}$ color wave functions,
where the nonrelativistic single quark spin wave function is
replaced by the relativistic solution $u_\alpha(\vec{x})$ of the
Dirac equation.

In our description of baryons we use the effective potential of
Eq.\ (\ref{potential}), which is given by a sum of a scalar
potential $S(r)$ providing confinement and the time component of a
vector potential $\gamma^{0}V(r)$. Obviously, other possible
Lorenz structures (e.g., pseudoscalar or axial) are excluded by
symmetry principles. It is known from lattice simulations that a
scalar potential should be a linearly rising one and the vector
potential is thought to be responsible for short-range
fluctuations of the gluon field
configurations~\cite{Takahashi:2000te}. In our study we
approximate $V_{\rm eff}(r)$ by a relativistic harmonic oscillator
potential with a quadratic radial
dependence~\cite{Lyubovitskij:2001nm}
\begin{eqnarray}\label{V_hop}
S(r) = M_1 + c_1 r^2\,, \hspace*{.5cm}
V(r) = M_2 + c_2 r^2\,.
\end{eqnarray}
The model potential defines unperturbed wave functions for the
quarks, whi\-ch are sub\-se\-quently used to calculate baryon
properties. This potential has no direct connection to the
underlying physical picture and is thought to serve as an
approximation of a realistic potential. Notice that this type of
potential was extensively used in chiral potential
models~\cite{Tegen:1983gg}-\cite{Gutsche:1989vy}. A positive
feature of this potential is that most of the calculations can be
done analytically. As was shown in
Refs.~\cite{Tegen:1983gg}-\cite{Gutsche:1989vy} and later on also
checked in the PCQM~\cite{Lyubovitskij:2001nm}, this effective
potential gives a reasonable description of baryon properties and
can be treated as a phenomenological approximation of the more
fundamental forces dictated by QCD.

The use of a variational {\it Gaussian ansatz} for the
effective potential (\ref{V_hop}) gives
the following solution for the ground state (for the excited
quark states we proceed by analogy):
\begin{eqnarray}
u_0(\vec{x}) \, = \, N \, \exp\biggl[-\frac{\vec{x}^{\, 2}}{2R^2}\biggr]
\, \left(
\begin{array}{c}
1\\ i \rho \, \frac{\vec{\sigma}\vec{x}}{R}\\
\end{array}
\right) \, \chi_s \, \chi_f\, \chi_c \, ,\label{wavefunction1}
\end{eqnarray}
where $N=[\pi^{3/2} R^3 (1+3\rho^2/2)]^{-1/2}$ is a normalization
constant, and $\chi_s$, $\chi_f$, $\chi_c$ are the spin, flavor
and color quark wave functions, respectively. The parameter
$\rho$, setting the strength of the "small component", can be
related to the axial charge $g_A$ of the nucleon. In the leading
order (3-quark-core) approximation, this relation
is~\cite{Lyubovitskij:2001nm}
\begin{equation}
g_A=\frac{5}{3}\biggl(1 - \frac{2\rho^2}{1+\frac{3}{2}\rho^2}\biggr) \,.
\end{equation}
The parameters of the effective potential $V_{\rm eff}$ can also be
expressed in terms of $\rho$ and~$R$:
\begin{equation}
M_1 = \frac{1 \, - \, 3\rho^2}{2 \, \rho R}, \;
M_2 = {\cal E}_0 - \frac{1 \, + \, 3\rho^2}{2 \, \rho R} , \;
c_1 \equiv c_2 =  \frac{\rho}{2R^3} \,.
\end{equation}
Here, ${\cal E}_0$ is the single-quark ground-state energy. In our
calculations we use the value $g_A$=1.25. Therefore, we have only
one free parameter in the model, $R$. In our numerical study, $R$
is varied in the region from 0.55 fm to 0.65 fm, which is set and
constrained by nucleon phenomenology~\cite{Lyubovitskij:2001nm}.
Such a variation of the parameter $R$ slightly changes the
physical quantities up to 5\%~\cite{Lyubovitskij:2001nm}. In this
paper we also test the sensitivity of the neutron EDM to a
variation of $R$.

The expectation value of an operator $\hat A$ is defined as
\begin{equation}\label{matrA}
\langle \hat A \rangle = ^B\!\!\langle\phi_0|\sum^{\infty}_{n=1}
\frac{i^n}{n!}\int d^4 x_1 \ldots \int d^4 x_n T[{\cal L}_I (x_1)
\ldots{\cal L}_I (x_n) \hat A] |\phi_0\rangle_c^B
\end{equation}
where ${\cal L}_I$ is the full interaction Lagrangian which may
contain both CP-even and CP-odd terms, as discussed below. The
superscript $``B"$ in Eq.~(\ref{matrA}) indicates that the matrix
elements are projected on the respective baryon states and the
subscript $``c"$ refers to contributions from connected graphs
only.

For the evaluation of Eq.~(\ref{matrA}) we apply Wick's theorem
with the appropriate propagators for quarks and mesons. For the
quark field we use a vacuum Feynman propagator for a fermion in
a binding potential. In the calculation of meson-quark loops we
include only the ground state in the quark propagator, which leads
to the following truncated form:
\begin{equation}
i G_q(x,y) = \langle 0 |T\{q(x)\bar q(y)\}| 0 \rangle \ \to \
\theta(x_0-y_0) u_0(\vec{x}) \bar u_0(\vec{y}) e^{-i{\cal E}_0
(x_0-y_0)}\,.
\end{equation}
Notice that in our previous papers~\cite{Lyubovitskij:2001nm} we
estimated explicitly the contribution of the low-lying excited
quark states in the quark propagator to the physical quantities.
Their contribution is about 10 to 15\% with respect to the ground
state contribution. Therefore, a restriction of the quark
propagator to the ground states is a reasonable  approximation.
For completeness we also calculate the corrections to the neutron
EDM due to the inclusion of excited quark states: the first
$p$-states ($1p_{1/2}$ and $1p_{3/2}$ in the non-relativistic
notation) and the second excited states ($1d_{3/2}, 1d_{5/2}$ and
$2s_{1/2}$), i.e. we restrict to the low-lying excited states with
energies smaller than the typical scale of $\Lambda =  1$ GeV of
low-energy approaches.

For the meson fields we use their free Feynman propagators:
\begin{equation}
i\Delta_{ij}(x-y)= \langle 0|T\{\Phi_i(x)\Phi_j(y)\}|0 \rangle
=\delta_{ij} \int\frac{d^4k}{(2\pi)^4i}
\frac{\exp[-ik(x-y)]}{M_\Phi^2-k^2-i\epsilon}.
\end{equation}

\section{The neutron electric dipole Form Factor and Moment in the PCQM}
\label{s3}

Here we apply the above described PCQM model to the calculation of
the electric dipole form factor (EDFF) and moment (EDM) of the
neutron. In this model, having the constituent quarks and
pseudoscalar mesons as basic degrees of freedom, the effective
chiral CP-odd Lagrangian at low energies can be written in the
following generic form \eq\label{EQS1} \hspace*{-2.25cm} {\cal
L}_{CPV}^{(1)}(x) &=& - \frac{i}{2}  \, \bar q(x) \, {\bf d_q} \,
\sigma^{\mu \nu} \, \gamma_5 \,  q(x) \, F_{\mu\nu}(x) \, + \, i
\, \bar q(x) \, e^{i\gamma_5 \hat\Phi(x)/(2F)} \, \gamma_5 \, {\bf
h_q} \, e^{i\gamma_5 \hat\Phi(x)/(2F)} \, q(x) \nonumber\\
\hspace*{-2.25cm}&=& - \frac{i}{2}  \, \bar q(x) \, {\bf d_q} \,
\sigma^{\mu \nu} \, \gamma_5 \,  q(x) \, F_{\mu\nu}(x) \, + \, i
\, \bar q(x) \, \gamma_5 \, {\bf h_q} \,  q(x) \nonumber\\
\hspace*{-2.25cm} &-& \frac{1}{2 F} \, \bar q(x) \,\, \{ {\bf h_q}
\, , \hat\Phi(x) \} \,\,  q(x) + O(\hat\Phi^2) \,,
\end{eqnarray}
where $F_{\mu \nu}$ is the electromagnetic field strength tensor,
the symbol $\{~,~\}$ denotes anticommutator, ${\bf d_q} = {\rm
diag}(d_u, d_d, d_s ) $ and ${\bf h_q} = {\rm diag} (h_u, h_d, h_s
) $ are diagonal matrices of the effective EDMs of the constituent
quarks and their CP-violating couplings to the chiral fields,
respectively. In Eq.~(\ref{EQS1}) we expanded the quark-meson term
in powers of $\hat\Phi/F$ and kept the leading non-trivial term
linear in $\hat\Phi$. It turns out that the first term in this
expansion does not contribute to the EDFF at one loop because the
diagrams involving this vertex do not contain the required
spin-flip structure $\vec\sigma_N \cdot \vec q$, the product of
the neutron spin operator $\vec\sigma_N$ and of the 3-momentum of
the photon $\vec q$. Therefore, in the following we keep only the
linear term in $\hat\Phi$ in Eq.~(\ref{EQS1}).

We assume that the quark-meson Lagrangian (\ref{EQS1}) originates
from a more fun\-da\-men\-tal level of quark and gluon CP-odd
interactions, so that the effective parameters  ${\bf d_q}$ and
${\bf h_q}$ can be related to those more fundamental CP-odd
parameters. We discuss these questions in section \ref{s4}.

Now, we calculate the neutron EDFF, $D_n(Q^2)$, starting from the
Lagrangian (\ref{EQS1}). The EDFF is defined in the standard way
through the neutron matrix element of the electromagnetic current
as \eq\label{Dn-def} \hspace*{-2cm} \langle n(p^\prime)| \,
J^{\mu}(0) \, |n(p)\rangle &=& \bar{u}_n(p^{\prime}) \biggl[
\gamma^\mu \, F_n^{(1)}(Q^2) \, + \, \frac{i}{2 m_N} \,
\sigma^{\mu\nu} \, q_{\nu} \, F_n^{(2)}(Q^2) \\ \hspace*{-2cm} &-&
\sigma^{\mu\nu} \, \gamma_5 \, q_{\nu} \, \, D_n(Q^2) \, + \,
(\gamma^\mu \, q^2 \, - \, 2 \, m_N \, q^\mu) \, \gamma_5 \,
A_n(Q^2) \biggr] u_n(p) \,, \nonumber
\en
where, in addition to $D_n(Q^2)$, $F_n^{(1)}(Q^2)$ and
$F_n^{(2)}(Q^2)$ are the well-known $CP$-even neutron
electromagnetic form factors and $A_n(Q^2)$ is the neutron anapole
moment form factor. The neutron EDM, $d_n$, is defined as the
value of the neutron EDFF at zero recoil, namely $d_n \, = \,
D_n(0)$.

To guarantee electromagnetic gauge invariance in non-covariant
approaches (see discussion in
Refs.~\cite{Miller:1997jr,Lyubovitskij:2001nm}) we have to work in
the Breit frame, which defines the momenta of the transferred
photon and the initial and final neutron states respectively as
\begin{equation}
q = \left( 0, \vec{q}\, \right) ,\; \; p = \left( E, - \vec{q}/2
\right) , \; \; p^\prime = \left(  E, \vec{q}/2 \right) ,
\end{equation}
where $E = \sqrt{m_N^2 + \vec{q}^{\,\,2}/4}$ is the nucleon
energy, $m_N$ the nucleon mass, and $q^2 \equiv - Q^2 = -
\vec{q}^{\,\,2}$ the momentum transfer squared. We can now use
Eq.~(\ref{Dn-def}) in the Breit frame and our model w.f. of
Eq.~(\ref{wavefunction1}) to extract the form factor $D_n (Q^2)$
as the term in $\langle J^{\, 0}(0) \rangle $ proportional to
$\vec{\sigma}_N \cdot \vec{q}$, namely
\begin{eqnarray}\label{Dn}
 \langle J^{\, 0}(0) \rangle\ =\
\frac{E}{m_N} \, \chi^\dagger_{_{N_{s^\prime}}} \ i \,
\vec{\sigma}_N \cdot \vec{q} \,\, \chi_{_{N_s}} \ D_n(Q^2) +
\ldots
 \nonumber .
\end{eqnarray}
Here $\chi_{N_s}$ and $\chi^\dagger_{N_{s^\prime}}$  are the
nucleon spin w.f.\ in the initial and final state, and
\eq\label{matrJ}
\hspace*{-2cm}
\langle J^0(0) \rangle \ =\ \langle n(p')|\sum^{\infty}_{n=1}
\frac{i^n}{n!}\int d^4 x_1 \ldots \int d^4 x_n T[{\cal L}_I (x_1)
\ldots{\cal L}_I (x_n) J^0(0)] |n(p)\rangle_c \ ,
\en
where ${\cal L}_I$ is the full interaction Lagrangian and the
subscript ``c" refers to connected graphs only.
In PCQM, the electromagnetic current operator \eq J^{\, \mu}(x) =
J_{\rm \Phi}^{\, \mu}(x) + J_{CPV}^{\ \mu}(x)
\en
contains a CP-conserving electromagnetic part, given in terms of
the charged pse\-u\-do\-sca\-lar meson fields:
\eq
J_{\rm \Phi}^{\mu}(x) \, = \, e \, \biggl[
\, \pi^-(x) \, i\partial^\mu\pi^+(x)
\, + \, K^-(x) \, i\partial^\mu K^+(x) \, \biggr] \ + \ {\rm h.c.}
\en
and a CP-violating part involving the quarks:
\begin{eqnarray}\label{EQS3}
J_{CPV}^{\ \mu}(x) \, = \, i  \, \partial_\nu \biggl[ \bar q(x) \,
{\bf d_q} \, \sigma^{\mu \nu} \, \gamma_5 \,  q(x) \biggr]
\,,\label{CPcurrent}
\end{eqnarray}
which is derived from the Lagrangian (\ref{EQS1}).
The interaction Lagrangian ${\cal L}_I$ in Eq.~(\ref{matrJ}) up to one
loop order is the linearized strong interaction Lagrangian between
quark and chiral fields, which consists of the CP-conserving
interaction given in Eq.~(\ref{L_eff_1}) and the CP-violating
Lagrangian of Eq.~(\ref{EQS1}):
\begin{equation}
{\cal L}_I(x) = - \bar q(x) i\gamma^5 \frac{\hat\Phi(x)}{F} S(r)
q(x) \, - \, \frac{1}{2F} \, \bar q(x) \,\, \{ {\bf h_q}  ,
\hat\Phi(x) \} \,\,  q(x) .\label{hterm}
\end{equation}
Up to one loop, the neutron EDFF  receives contributions from
three types of effective diagrams, shown in Fig.~1: the
``$h$-terms" (Fig.~1a), which are proportional to the CP-violating
couplings $h_q$ of Eq.~(\ref{hterm}), the ``tree-level terms"
(Fig.~1b) and the ``one-loop corrections" (Fig.~1c), both due to
the CP-violating current of Eq.~(\ref{CPcurrent}) and proportional
to the couplings $d_q$. The diagrams in Figs.~1a and 1c encode the
chiral corrections to the neutron EDM. The diagram in Fig.~1a has
contributions from charged pion and kaon clouds, while the diagram
in Fig.~1c contains contributions from $\pi$, $K$ and $\eta$-meson
clouds.

The total contribution of Fig.~1a to the neutron EDFF is then
composed by charged pion and Kaon clouds as:
\begin{eqnarray} \label{Eq_1}
D_n^{[h]}(Q^2) \, = \, \sum\limits_{\Phi = \pi, K} D_n^{[h;
\Phi]}(Q^2) \,, \ \ \ \ \ \ \ D_n^{[h; \Phi]}(Q^2) \, \equiv \,
d^{[h; \Phi]} \, F^{[h;\Phi]}(Q^2)
\end{eqnarray}
where the normalization $F^{[h;\Phi]}(0)=1$ defines $d^{[h;
\Phi]}$ as the contributions of the charged $\pi$ and $K$ clouds
of Fig.1a to the neutron EDM:
\begin{eqnarray}\label{Eq_3}
d_n^{[h; \Phi]} \equiv D_n^{[h; \Phi]}(0) \, = \, e \,
c^{[h;\Phi]} \, \frac{g_{\pi NN} \bar g_{\Phi NN}}{2 m_N}
\int\frac{d^3k}{(2\pi)^3} \, \frac{F_{\pi NN}(\vec{k}^{\,2}) \,
\bar F_{\pi NN}(\vec{k}^{\,2})}{w^3_\Phi(\vec{k}\,)}
\end{eqnarray}
and $F^{[h;\Phi]}(Q^2)$, the form factor normalized to unity at
zero recoil is given by:
\begin{eqnarray}
\hspace*{-.25cm}
F^{[h;\Phi]}(Q^2) &=& \frac{I^{[h;\Phi]}(Q^2)}{I^{[h;\Phi]}(0)}\,,\\ 
\hspace*{-.25cm} 
I^{[h;\Phi]}(Q^2) &=& \frac{m_N}{E} \, \int\frac{d^3k}{(2\pi)^3} \,
\frac{F_{\pi NN}[(\vec{k} + \vec{q}\,)^2] \, \bar F_{\pi
NN}(\vec{k}^{\,2})}{w_\Phi(\vec{k}+\vec{q}\,) \, w_\Phi(\vec{k})}
\, \frac{2}{w_\Phi(\vec{k} + \vec{q} \,) \,+ \, w_\Phi(\vec{k}\,)}
\nonumber
\end{eqnarray}
Here $\Phi = \pi$ or $K$, $w_\Phi(\vec{q}\,) = \sqrt{M_\Phi^2 +
\vec{q}^{\,\,2}}$ is the meson energy, and $c^{[h;\Phi]}$ is the
SU(6) spin-flavor factor, which is $c^{[h;\pi]} = 1/2$ for the
pion-loop and $c^{[h;K]} = 1/5$ for the kaon-loop diagram.

The expression for $d_n^{[h;\Phi]}$ in Eq.~(\ref{Eq_3}) is written
in terms of the known $\pi NN$ coupling which satisfies the
Goldberger-Treiman relation
 \eq
 g_{\pi NN} = g_A \, \frac{m_N}{F},
 \en
(where $g_A=1.25$ is the axial nucleon charge) and the
CP-violating couplings $\bar g_{\Phi NN}$, which are:
 \eq
 \bar g_{\pi NN} = \frac{h_u + h_d}{2 F} \, \gamma \,, \hspace*{1cm}
 \bar g_{KNN} = \frac{h_u + h_s}{2 F} \, \gamma .
 \en
Here $\gamma$ is the isovector-scalar two-quark condensate in the
nucleon, given by:
\begin{equation}
\langle N|\bar q \tau_3 q|N \rangle \, = \,
\gamma \, \bar u_N \tau_3 u_N \, .
\end{equation}
This factor $\gamma$ coincides with the so-called relativistic
reduction factor~\cite{Lyubovitskij:2001nm}. In the PCQM, $\gamma
= 5/8$~\cite{Lyubovitskij:2001nm}. Finally, $F_{\pi NN}$ and $\bar
F_{\pi NN}$ are the normalized CP-conserving and CP-violating form
factors that appear in PCQM \cite{Lyubovitskij:2001nm}, and which
regularize the divergent loop integral:
 \eq F_{\pi NN}(\vec{k}^{\,2}) \, = \, 
\exp( - \vec{k}^{\,2} R^2/4) \, \biggl[
 1 + \frac{\vec{k}^{\,2} R^2}{8} \biggl( 1 - \frac{5}{3 g_A}
 \biggr)\biggr] \,,
 \en
and
 \eq
\bar F_{\pi NN}(\vec{k}^{\,2}) \, = \, \exp( - \vec{k}^{\,2}
R^2/4) \, \biggl[ 1 + \frac{\vec{k}^{\,2} R^2}{4} \frac{1 -
3g_A/5}{9g_A/5 - 1} \biggr]
 \en
Now, it is worth checking if our result for the pion-cloud
contribution to $d_n^{[h; \pi]}$ in Eq.~(\ref{Eq_3}) is consistent
with the model-independent prediction derived in
Ref.~\cite{Crewther:1979pi} for the leading term in the chiral
expansion:
 \eq\label{Eq_4}
 \bar d_n^{[h; \pi]} \, = \, e \,
\frac{g_{\pi NN} \, \bar g_{\pi NN}}{4 \, \pi^2 \, m_N} \, {\rm
log}\frac{m_N}{M_\pi} \,.
 \en
To this end, we drop the normalized form factors $F_{\pi NN}$ and
$\bar F_{\pi NN}$ in Eq.~(\ref{Eq_3}) by substituting $F_{\pi
NN}=\bar F_{\pi NN}=1$, and analyze this equation using
alternatively cutoff or dimensional regularizations. Both methods
of regularization give the same result, which also coincide with
Eq.~(\ref{Eq_4}). Therefore, our approach is consistent with QCD
in the local limit, when $F_{\pi NN} = \bar F_{\pi NN} =1$. The
nontrivial form factors $F_{\pi NN}$ and $\bar F_{\pi NN}$ provide
an ultraviolet convergence for the EDM. Notice that
Eq.~(\ref{Eq_4}) has been derived in ChPT when the strong
CP-violating $\pi NN$ coupling constant $\bar g_{\pi NN}$ is
defined by the $\theta$-term. This result is also valid for
general CP-violating couplings. The difference is encoded in the
redefinition of \eq \bar g_{\pi NN}(\bar\theta) \, = \, \bar\theta
\, \frac{\bar m}{F} \, \gamma \, \to \, \bar g_{\pi NN}(h_u,h_d)
\, = \, \frac{h_u + h_d}{2 F} \, \gamma \,.
\en
The leading contribution of the kaon-cloud diagram is also
proportional to the chiral logarithm but contains a
model-dependent coefficient $c^{[h;K]} = 1/5$:
\begin{equation}\label{Eq_4_K}
\bar d_n^{[h; K]} \, = \, c^{[h;K]} \, e \, \frac{g_{\pi NN} \,
\bar g_{\pi NN}}{4 \, \pi^2 \, m_N} \, {\rm log}\frac{m_N}{M_K}
\,.
\end{equation}
We remark that the coefficient $c^{[h;K]}$ was calculated
previously in Heavy Baryon Chiral Perturbation
Theory~(HBChPT)~\cite{Borasoy:2000pq}, where it was expressed
through the parameters of the chiral Lagrangian as:
\begin{eqnarray}
c^{[h;K]} \, = \, \frac{D-F}{D+F} \, \frac{b_F-b_D}{b_F+b_D} .
\end{eqnarray}
Here $D$ and $F$ are the axial-vector couplings, and $b_D$ and
$b_F$ are low-energy constants. Using the actual
values~\cite{Borasoy:2000pq} $D = 0.80$ and $F = 0.46$ fixed from
a fit of semileptonic hyperon decays, and $b_D = 0.079$
GeV$^{-1}$, $b_F = - 0.316$ GeV$^{-1}$ determined from the
calculation of baryon masses and the $\pi$-$N$ sigma-term up to
fourth order in the chiral expansion, we deduce the prediction of
HBChPT for $c^{[h;K]} = 0.45$. This is more than a factor two
larger than the prediction of our model.

Now, let us examine the diagram Fig.~1b.  In this calculation we
use the well-known quark model formula for the projection of the
spin-flavor part of the Lagrangian~(\ref{EQS1}) between neutron
states:
\begin{equation}
\langle n \uparrow \vert \sum_{i=1}^3 \, ({\bf d}_q)_i \,
(\vec{\sigma} \cdot \hat{q}\,)_i \; \vert n \uparrow \rangle =
\frac{4}{3} d_d - \frac{1}{3} d_u \,,
\end{equation}
where the sum is over the constituents,  ${\bf d}_q$ is a diagonal
$2 \times 2$ flavor matrix of the quark EDMs, $\hat{q} =
\vec{q}/|\vec{q}\, |$ is the unit vector along the momentum
$\vec{q}$, and $\vec\sigma$ are the Pauli matrices in spin space.

Then the contribution of the diagram in Fig.~1b (a tree-level
contribution) to the neutron EDFF is given by
\begin{eqnarray}
& &D_n^{[d; tree]}(Q^2) = \left( \frac{4}{3} d_d - \frac{1}{3} d_u
\right) \; \left( \frac{1}{2} + \frac{3}{10} g_A \right) \;
F^{[d]}(Q^2) \,, \label{treelevel}\\
 & & F^{[d]}(Q^2) = \frac{m_N}{E}
\; \exp(- Q^2 \, R^2/4) \; \left[ 1 + \frac{Q^2 R^2}{4} \frac{1 -
3g_A/5}{1 + 3g_A/5} \right]\,, \nonumber
\end{eqnarray}
where $F^{[d]}(0) = 1$.

Finally, we present the loop correction diagram of Fig.~1c. This
one has three contributions: from $\pi$, $K$ and $\eta$ loops,
which we denote as:
\begin{eqnarray}
\hspace*{-1cm} D_n^{[d; loop]}(Q^2) &=& D_n^{[d; \pi]}(Q^2) +
D_n^{[d; K]}(Q^2) + D_n^{[d; \eta]}(Q^2) \,, \label{loops}
\end{eqnarray}
and where each meson loop contribution is given by:
\begin{eqnarray}
\hspace*{-.5cm}
D_n^{[d; \Phi]}(Q^2) = - c^{[d;\Phi]} \, \frac{g_{\pi NN}^2}{200
\,\, m_N^2} \,  F^{[d]}(Q^2) \, \int\frac{d^3k}{(2\pi)^3} \,
\vec{k}^{\,2} \, \frac{F^2_{\pi
NN}(\vec{k}^{\,2})}{w^3_\Phi(\vec{k}\,)}  ,\quad  \Phi = \pi, K,
\eta . \nonumber
\end{eqnarray}
Here $c^{[d;\Phi]}$ are SU(6) spin-flavor factors: $c^{[d;\pi]} =
7 d_u + 2 d_d$\,, $c^{[d;K]} = 6 d_s$  and $c^{[d;\eta]} = 4d_u/3
- d_d/3$.

The total neutron EDFF, up to one loop in the chiral expansion, is
then given by the tree level contributions of Fig.~1.b and the
loop contributions of Figs.~1.a and 1.c:
\begin{eqnarray}\label{EDFF_tree+loop}
\hspace*{-1cm} D_n(Q^2) &=& D_n^{[tree]}(Q^2) \, + \,
D_n^{[loop]}(Q^2) \,,\nonumber \\[3mm] \hspace*{-1cm}
D_n^{[tree]}(Q^2) &\equiv& D_n^{[d; tree]}(Q^2) \,, \\[3mm]
\hspace*{-1cm} D_n^{[loop]}(Q^2) &\equiv& \sum_{\Phi = \pi, K}
D_n^{[h; \Phi]}(Q^2) \, + \sum_{\Phi = \pi, K, \eta} D_n^{[d;
\Phi]}(Q^2) \,, \nonumber
\end{eqnarray}
where
\begin{eqnarray}
& & D_n^{[d; tree]}(Q^2) = a^{[tree]} \left( \frac{4}{3} d_d -
\frac{1}{3} d_u \right) \, F^{[d]}(Q^2) \,,\nonumber\\
& & D_n^{[d; \pi]}(Q^2)  = a^{[\pi]} \left( \frac{7}{9} d_u +
\frac{2}{9} d_d \right) \, F^{[d]}(Q^2) \,, \nonumber\\
& &D_n^{[d; K]}(Q^2)    = a^{[K]} \ d_s\ F^{[d]}(Q^2) \,,\nonumber\\[2mm] 
& & D_n^{[d; \eta]}(Q^2) = a^{[\eta]} \left(
\frac{4}{3} d_u - \frac{1}{3} d_d \right) \, F^{[d]}(Q^2) \,,\\
& & D_n^{[h; \pi]}(Q^2)  = \frac{e}{\Lambda_\chi^2} \,
b^{[\pi]} \, \left( \frac{1}{2} h_u + \frac{1}{2} h_d \right) \,
F^{[h; \pi]}(Q^2) \,, \nonumber\\
& & D_n^{[h; K]}(Q^2)  =
\frac{e}{\Lambda_\chi^2} \, b^{[K]} \, \left( \frac{1}{2} h_u +
\frac{1}{2} h_s \right) \, F^{[h; K]}(Q^2) \,, \nonumber
\end{eqnarray}
where in the contributions of the diagram in Fig.~1a, for
convenience, we use the dimensional parameter, $\Lambda_\chi = 4
\pi F_\pi \simeq 1.2$ GeV, the scale of spontaneously broken
chiral symmetry, where $F_\pi = 92.4$ MeV. The corresponding
numerical coefficients  $a^{[tree]} = 7/8$, $a^{[\pi]} = 3 \times
10^{-2}$, $a^{[K]}   = 6.3 \times 10^{-3}$, $a^{[\eta]} = 8.5
\times 10^{-4}$, $b^{[\pi]} = 2.92$ and $b^{[K]} = 0.10$ are
obtained from the evaluation of Eqs.~(\ref{Eq_3}),
(\ref{treelevel}) and (\ref{loops}).

It is convenient to separate the contributions from the individual
quarks as:
\begin{equation} 
\hspace*{-.5cm}
D_n(Q^2)  \equiv \sum_{q = u, d, s} D_{n,q}(Q^2) \, =
\sum\limits_{q = u, d, s} \, \biggl[ \, \Delta_q^{[d]} \, d_q \,
F^{[d]}(Q^2) \, + \, \Delta_q^{[h]} \, \tilde h_q \, F^{[h]}_q
(Q^2) \, \biggr]   ,\label{Eq_sep}
\end{equation}
where we define $\tilde h_q = (e/\Lambda_\chi^2) \, h_q$ and
\begin{eqnarray}
F^{[h]}_u(Q^2)&=&\frac{b^{[\pi]} \, F^{[h;\pi]}(Q^2) \, + \,
b^{[K]} \, F^{[h; K]}(Q^2)}{b^{[\pi]} \, + \, b^{[K]}} \,,
\nonumber
\\ [2mm] F^{[h]}_d(Q^2) &=& F^{[h;\pi]}(Q^2) \,,\\[5mm] F^{[h]}_s(Q^2) &=&
F^{[h; K]}(Q^2). \nonumber
\end{eqnarray}
The coefficients $\Delta_q^{[d]}$ and  $\Delta_q^{[h]}$ are:
\begin{eqnarray}\label{delta-d-h}
\Delta_u^{[d]} &=& - \frac{1}{3} a^{[tree]}
             - \frac{7}{9} a^{[\pi]}
             - \frac{4}{3} a^{[\eta]} =  - 0.316 \,,
\nonumber\\ \Delta_d^{[d]} &=&   \frac{4}{3} a^{[tree]}
             - \frac{2}{9} a^{[\pi]}
             + \frac{1}{3} a^{[\eta]} =  1.16\,, \nonumber\\
\Delta_s^{[d]} &=& - a^{[K]} =  -0.006\,, \label{Deltas} \\
\Delta_u^{[h]} &=& \frac{1}{2} b^{[\pi]} \, + \, \frac{1}{2}
b^{[K]} = 1.51 \,, \nonumber\\ \Delta_d^{[h]} &=& \frac{1}{2}
b^{[\pi]} = 1.46 \,, \nonumber\\ \Delta_s^{[h]} &=& \frac{1}{2}
b^{[K]} = 0.05 \,. \nonumber
\end{eqnarray}
The numerical results for the $\theta$-term contribution to the
neutron EDFF (including the pion and kaon loops) are shown in
Fig.~2. In Fig.~3 we plot the $Q^2$-dependence of the $d$ quark
contributions to the neutron EDFF arising from its individual EDM,
$d_d$, as well as its CP-violating coupling to mesons, $h_d$,
contained in the first and second terms of Eq.~(\ref{Eq_sep}),
respectively. Since the values of $d_d$ and $\tilde h_q$ and their
relative signs are not known, we display these two types of
contributions separately, in terms of normalized form factors:
$F_{n,d}^{[d]}(Q^2) \, = \, \Delta_d^{[d]} \, F^{[d]}(Q^2)$ for
the individual $d$ quark EDM contribution, and $F_{n,d}^{[h]}(Q^2)
\, = \, \Delta_d^{[h]} \, F^{[h]}_d(Q^2)$ for the $d$ quark
contribution via its CP-violating coupling to mesons.

The neutron EDM, $d_n$, is then obtained by definition as the
neutron EDFF at zero recoil. Using $F^{[d]}(0)=1$ and
$F^{[h]}_q(0)=1$, for $q=u, d, s$, the resulting EDM  can also be
expressed in terms of the individual quarks EDMs $d_q$ and
CP-violating couplings $\tilde h_q$, as: \eq\label{tree+loop} d_n
\, = \, \sum\limits_{q=u,d,s} \, [ \, \Delta_q^{[d]} \, d_q \, +
\, \Delta_q^{[h]} \, \tilde h_q \, ] \,,
\en
with the coefficients $\Delta_q^{[d]}$ and $\Delta_q^{[h]}$ given
in Eq.~(\ref{Deltas}).

Another CP-odd parameter which may have important implications for
CP-violation in atoms is the electron-neutron Schiff moment
$S^{\prime}$, related to the neutron EDM form factor $D_n(Q^2)$
by~\cite{Thomas:1994wi}
 \eq\label{Shiff-def}
 S^{\prime} = -
\left[\frac{d D_n(Q^2)}{dQ^2}\right]_{Q^2=0}.
 \en 
From the expression (\ref{EDFF_tree+loop})-(\ref{delta-d-h}) we
obtain: \eq\label{Schiff-PCQM} S^{\prime} \, = \,  S^{\prime \,
[h]} + S^{\prime \, [d]} \,,
\en
where \eq\label{Schiff-PCQM1} \hspace*{-2cm} S^{\prime \, [h]} &=&
e \ \sum\limits_{\Phi = \pi, K} \ c^{[h; \Phi]} \ \frac{g_{\pi NN}
\ \bar g_{\Phi NN}}{2 m_N} \int\frac{d^3k}{(2\pi)^3} \,
\frac{F_{\pi NN}(\vec{k}^{\,2}) \, \bar F_{\pi
NN}(\vec{k}^{\,2})}{w^3_\Phi(\vec{k}\,)} \nonumber\\[3mm]
\hspace*{-1cm} &\times&\biggl[ \frac{3}{4 \, w^2_\Phi(\vec{k}\,)}
-
 \frac{5}{6} \, \frac{\vec{k}^{\,2}}{w^4_\Phi(\vec{k}\,)}
+ \frac{1}{8 \, m_N^2} - \frac{M_\Phi^2}{w^2_\Phi(\vec{k}\,)} \,
\frac{F_{\pi NN}^{\prime}(\vec{k}^{\,2})}{F_{\pi NN}(\vec{k}^{\,2})}
- \frac{2}{3} \, \vec{k}^{\,2} \,
\frac{F_{\pi NN}^{\prime\prime}(\vec{k}^{\,2})}{F_{\pi NN}(\vec{k}^{\,2})}
\biggr]\,   \nonumber
\en
is the contribution of the diagram in Fig.~1a and
\eq\label{Schiff-PCQM2} 
\hspace*{-2.25cm} S^{\prime \, [d]} &=& \left
[ \left( \frac{4}{3} d_d - \frac{1}{3} d_u \right) \; \left(
\frac{1}{2} + \frac{3}{10} g_A \right) - \sum\limits_{\Phi = \pi,
K, \eta} \, c^{[d;\Phi]} \, \frac{g_{\pi NN}^2}{200 \, \, m_N^2}
\, \int\frac{d^3k}{(2\pi)^3} \, \vec{k}^{\,2} \, \frac{F^2_{\pi
NN}(\vec{k}^{\,2})}{w^3_\Phi(\vec{k}\,)} \right] \nonumber\\
\hspace*{-2.25cm} &\times&\left[ \frac{1}{8 m_N^2} \, + \,
\frac{3}{2} \, R^2 \, \frac{g_A}{1 \, + \, 3 g_A/5} \right] \equiv
\sum\limits_{q = u,d,s} \, \Delta_q^{[d]} \, d_q \, \left[
\frac{1}{8 m_N^2} \, + \, \frac{3}{2} \, R^2 \, \frac{g_A}{1 \, +
\, 3 g_A/5} \right]\, \nonumber
\en
is the contribution of the diagrams in Figs.~1b and 1c.
Here $F_{\pi NN}^{\prime}$ and $F_{\pi NN}^{\prime\prime}$
are the first- and second-order derivatives of the $F_{\pi NN}$ form
factor with respect to $\vec{k}^{\,2}$.

Again, as for the case of the EDM, we can show that our Schiff
moment is consistent with the ChPT results in leading order of the
chiral expansion. As in the case of the neutron EDM, we drop the
normalized form factors $F_{\pi NN}$ and $\bar F_{\pi NN}$ in
Eq.~(\ref{Schiff-PCQM}), substituting $F_{\pi NN}=\bar F_{\pi
NN}=1$ and, therefore, $F_{\pi NN}^{\prime} = F_{\pi
NN}^{\prime\prime} = 0$. Then we analyse the leading term in the
chiral expansion (which is ultraviolet convergent):
\begin{eqnarray}\label{Schiff-PCQM_LO}
\bar S^{\prime} \, = \, \, e \, \sum\limits_{\Phi = \pi, K} \,
c^{[h;\Phi]} \, \frac{g_{\pi NN} \bar g_{\Phi NN}}{2 m_N}
\int\frac{d^3k}{(2\pi)^3} \, \frac{1}{w^5_\Phi(\vec{k}\,)} \,
\biggl[ \frac{3}{4} -
 \frac{5}{6} \, \frac{\vec{k}^{\,2}}{w^2_\Phi(\vec{k}\,)} \biggr]\,.
\end{eqnarray}
A straightforward calculation of the integral (\ref{Schiff-PCQM_LO})
gives
\begin{eqnarray}\label{Schiff-PCQM_LO_2}
\bar S^{\prime} = \, e \, \sum\limits_{\Phi = \pi, K} \,
c^{[h;\Phi]} \, \frac{g_{\pi NN}\,\bar g_{\Phi NN}}{48 \, \pi^2 \,
m_N \, M_\Phi^2} \,.
\end{eqnarray}
In the case when we restrict  Eq.~(\ref{Schiff-PCQM_LO_2}) to the
contribution of the pion-cloud, our result coincides with the
leading order result of ChPT~\cite{Thomas:1994wi,Hockings:2005cn}:
\begin{eqnarray}\label{Schiff_LO}
\bar S^{\prime} =
e  \, \frac{g_{\pi NN}\,\bar g_{\pi NN}}{48 \, \pi^2 \, m_N \, M_\pi^2}\,.
\end{eqnarray}
Let us now present our results for the electron-neutron Schiff
moment $S^{\prime}$ in terms of the partial contributions:

{\bf(a)} Diagram in Fig.~1a: partial $\pi$ and $K$ meson loop
contributions
 \eq\label{Schiff_h_piK} \hspace*{-2cm} S^{\prime \,
[d; \pi]} = 16.22 \ {\rm GeV}^{-2} \times \biggl[ \frac{1}{2}
\tilde h_u + \frac{1}{2} \tilde h_d \biggr], \ \ \ S^{\prime \,
[h; K]} &=& 0.21 \ {\rm GeV}^{-2} \times \biggl[ \frac{1}{2}
\tilde h_u + \frac{1}{2} \tilde h_s \biggr] \,.
\en

{\bf(b)} Diagram in Fig.~1b: tree-level result
 \eq\label{Schiff_d_tree} \hspace*{-2cm} S^{\prime \, [d; tree]} =
8.79 \ {\rm GeV}^{-2} \times \biggl[ \frac{4}{3} d_d - \frac{1}{3}
d_u \biggr] \,.
\en

{\bf(c)} Diagram in Fig.~1c: partial $\pi$, $K$ and $\eta$ meson
loop contributions
 \eq\label{Schiff_d_loop} \hspace*{-2cm}
S^{\prime \, [d; \pi]} &=& - 0.31  \ {\rm GeV}^{-2} \times \biggl[
\frac{7}{9} d_u + \frac{2}{9} d_d \biggr] \,, \nonumber\\[3mm]
\hspace*{-2cm} S^{\prime \, [d; K]} &=& - 0.06 \ {\rm GeV}^{-2}
\times d_s \,, \\[3mm] \hspace*{-2cm} S^{\prime \, [d; \eta]} &=&
- 0.01 \ {\rm GeV}^{-2} \times \biggl[ \frac{4}{3} d_u -
\frac{1}{3} d_d \biggr] \,. \nonumber
\en
Next we summarize our result for the Schiff moment in terms of the
individual quarks EDMs $d_q$ and CP-violating couplings $\tilde
h_q$, as:
 \eq\label{tree+loop_D} \hspace*{-2cm} S^{\prime} =
\sum\limits_{q=u,d,s} \, [ \beta^{[d]} \, \Delta_q^{[d]} \, d_q \,
+ \, \beta_q^{[h]} \, \Delta_q^{[h]} \, \tilde h_q \, ] \,,
 \en
with the coefficients $\Delta_q^{[d]}$ and $\Delta_q^{[h]}$ given
in Eq.~(\ref{Deltas}). Here, for convenience we introduced the
$\beta$-factors: $\beta^{[d]} = - F^{[d]\, \prime}(0)$ and
$\beta_q^{[h]} = - F_q^{[h] \, \prime}(0)$. Their numerical values
are: \eq \hspace*{-2.5cm} \beta^{[d]} = 10.06 \ {\rm GeV}^{-2},
\ \ \beta_u^{[h]} = 5.44 \ {\rm GeV}^{-2}, \ \ \beta_d^{[h]} =
5.55 \ {\rm GeV}^{-2}, \ \ \beta_s^{[h]} = 2.10 \ {\rm
GeV}^{-2}.
\en

\section{Limits on the quark EDMs and their phenomenological implications}
\label{s4}

Here, using Eq. (\ref{tree+loop}) we extract the limits on the
quark EDMs, $d_q$, and CP-violating couplings $h_q$, from the
existing experimental upper limit on the neutron EDM, shown in
Eq.~(\ref{EDM-exp-1}). Assuming the absence of accidental
cancellations between the different terms  in
Eq.~(\ref{tree+loop}) we obtained the corresponding limits shown
in the Table~1.

In order to discuss possible phenomenological implications of
these limits and compare them with the results  of other
approaches, it would be desirable to express the individual quark
EDMs in terms of more fundamental CP violating parameters
specifying the origin of CP violation.

The effective CP-odd Lagrangian in terms of the quark and
gluon fields up to operators of dimension~six has the following
standard form~\cite{Weinberg:1989dx}-\cite{Demir:2002gg}:
\eq\label{eff56}
\hspace*{-2cm}
{\cal L}_{\rm eff}^{\cpv}(x)&=& \frac{\bar\theta}{16 \pi^{2}} {\rm tr}
\big(\widetilde{G}_{\mu\nu} G^{\mu\nu} \big)
- \frac{i}{2} \, \bar q \,\, {\bf d_q^E} \,\,
\sigma^{\mu \nu} \, \gamma_5 \, F_{\mu\nu} \, q
- \frac{i}{2} \bar q \, \,\, {\bf d_q^C} \,\,
\sigma^{\mu \nu} \, \gamma_5 \,
G_{\mu\nu}^a \, T^a \, q \,\nonumber\\
\hspace*{-2cm}
&-&  \frac{1}{6} \, C_W \,
f^{abc} \, G^a_{\mu\alpha}  \, G^{b\alpha}_{\nu}
\, G^c_{\rho\sigma} \, \varepsilon^{\mu\nu\rho\sigma} \,,
\en
where $G^a_{\mu\nu}$ is the gluon stress tensor,
$\widetilde{G}_{\mu\nu}=\frac{1}{2}
\epsilon_{\mu\nu\sigma\rho}{G}^{\sigma \rho}$ is its dual tensor,
and  $T^a$ and $f^{abc}$ are the $SU(3)$ generators and structure
constants, respectively. In this equation the first term
represents the SM QCD $\theta$-term, while the last three terms
are the non-renormalizable effective operators induced by physics
beyond the SM. The second and third terms are the dimension-five
electric and chromoelectric dipole quark operators, respectively,
and the last term is the dimension-six Weinberg
operator~\cite{Weinberg:1989dx}. The quark EDMs, ${\bf d_q^E} =
{\rm diag} (d_u^E, d_d^E, d_s^E)$, and quark chromoelectric dipole
moments (CEDM), ${\bf d_q^C} = {\rm diag}(d_u^C, d_d^C, d_s^C)$,
form diagonal matrices in flavor space. The operators of the above
Lagrangian can be induced by physics beyond the SM after
integrating out the heavy degrees of freedom.

The contribution of the QCD $\theta$-term can be related to the
parameters of the Lagrangian (\ref{EQS1}) on the basis of the
following standard reasoning. A chiral $U(1)$ transformation in
flavor space allows one to remove the gluonic $\theta$-term from
the Lagrangian (\ref{eff56}) and pass it as a CP-violating complex
phase to the quark mass operators. For a small value of $\theta$,
the CP-violating term becomes (for details see
Refs.~\cite{Baluni:1978rf,Crewther:1979pi,Fujikawa:1979ay}):
\begin{equation}
{\cal L}_{CPV}^{str (0)} =  i \bar\theta \bar m  \, \sum_{q=u,d,s}
\bar{q}(x) \gamma_5 q(x),
\end{equation}
where $q(x)$ denotes a triplet in flavor space, and so this term
is a flavor-$SU(3)$ singlet. The mass coefficient is:
\begin{equation}
\bar{m} = \frac{m_u m_d m_s}{m_u m_d + m_u m_s + m_d m_s},
\end{equation}
which would vanish if any of the quarks were massless. From this
term one can construct an effective chiral Lagrangian by
introducing the chiral field $e^{i \gamma_5 \hat\Phi/F}$ and
expanding it in powers of $\hat\Phi/F$:
\begin{eqnarray}\label{LCP}
{\cal L}_{CPV}^{str} &=& i \bar\theta \bar m \bar q \gamma_5
\exp\left(i\gamma_5 \frac{\hat\Phi}{F} \right) q  =
 i \bar\theta \bar{m} \bar q \gamma_5 q - \bar\theta \bar{m} \bar q
\frac{\hat\Phi}{F} q + O(\hat\Phi^2) \,.
\end{eqnarray}
For the reasons we gave after Eq.~(\ref{EQS1}), the leading
contribution to the EDFF comes from the term linear in $\Phi$ in
the expansion of Eq.~(\ref{LCP}). Comparing the Lagrangians in
Eqs.~(\ref{EQS1}) and (\ref{LCP}) we obtain the contribution of
the $\theta$-term to the parameter $\bf{h}_q$ which is, naturally,
flavor-blind:
\begin{eqnarray}\label{h-theta}
{\bf h_q} \,\, = \,\, {\bf h^{BSM}_q} \,\, + \,\, \bar\theta \,\, \bar m
\,\, {\bf I_{3 \times 3}} \,,
\end{eqnarray}
where ${\bf I_{3 \times 3}}$ is the unit matrix in flavor space.
Here we have singled out the part ${\bf h^{BSM}_q}$ which does not
contain the $\theta$ contribution but is due to the other terms in
the Lagrangian~(\ref{eff56}).

In order to relate the parameters $d_q$ and $h^{BSM}_q$ of the
Lagrangian~(\ref{EQS1}) with the parameters $d_q^E$, $d_q^C$ and
$C_W$ of the Lagrangian (\ref{eff56}) we apply a simplified
approach relying on the so-called {\it na\"\i ve dimensional
analysis}~\cite{Weinberg:1989dx,Manohar:1983md,Georgi:1986kr}. In
this way one obtains the following order-of-magnitude relations:
\eq d_q &=&  d_q^E \, + \, \frac{e}{4\pi} \, d_q^C \, + \, \frac{e
\, \Lambda}{4\pi} \, C_W \,, \label{NDA1} \\ h^{BSM}_q &=& \,
\frac{\Lambda^2}{4\pi} \, d_q^C \, + \, \frac{\Lambda^3}{4\pi} \,
C_W \, . \label{NDA2}
\en
We assume all the operators in Eq.~(\ref{eff56}) to be normalized
at an energy scale of about $ 1$~GeV, where the perturbative
quark-gluon picture is still reasonable and, on the other hand,
occurs an overlap with the chiral symmetry breaking scale
$\Lambda_\chi = 4 \pi F_\pi \simeq 1.2$ GeV, which corresponds to
the normalization scale of the Lagrangian (\ref{EQS1}). Thus, all
the parameters in Eqs.~(\ref{NDA1}) and (\ref{NDA2}) are at the
same scale $\Lambda_\chi \simeq 1.2$~GeV. In such a case, it is
also natural to identify the parameter $\Lambda$ with
$\Lambda_\chi$. In a specific model beyond-SM, the parameters of
the effective Lagrangian (\ref{eff56}) could be calculable in
terms of the model parameters at some high energy scale. These
parameters then must be QCD-evolved down to the hadronic scale
$\Lambda_\chi$ corresponding to the scale of the neutron EDM. The
renormalization-group relations of the parameters $d_q^E$,
$d_q^C$, $C_W$ in Eqs.~(\ref{NDA1}) and (\ref{NDA2}) with their
values calculated at a high energy scale can be found in
Refs.~\cite{Arnowitt:1990eh,Degrassi:2005zd}.

Now, using Eqs.~(\ref{h-theta}), (\ref{NDA1}), (\ref{NDA2}) in
Eq.~(\ref{tree+loop}), we can express the neutron EDM in terms of
the more ``fundamental" parameters $\bar\theta$,  $d_q^E$, $d_q^C$
and  $C_W$ of the Lagrangian~(\ref{eff56}) as \eq\label{d-w} &
&d_n = \sum_{q=u,d,s}(z_q^E d_q^E + z_q^C d_q^C) \, + \, z_W \,
\Lambda_\chi \, C_W + (\bar\theta\ z_{\theta} \times  10^{-16}\ e
\cdot cm)\, .
\en
The numerical values of the $z$-coefficients are given in Table 2, 
where we also present for comparison the values of these 
coefficients derived in some other approaches:   
QCD sum rules~\cite{Pospelov:1999mv,Olive:2005ru},    
parton quark model~\cite{Abel:2004te},  
SU(6) model~\cite{He:1989xj}, MIT bag model~\cite{Baluni:1978rf}, 
current algebra~\cite{Crewther:1979pi}, effective chiral  
approach~\cite{Pich:1991fq}, HBChPT~\cite{Borasoy:2000pq},  
chiral bag model~\cite{Musakhanov:1984qy}, cloudy bag  
model~\cite{Morgan:1986yy}, chiral quark-meson  
model~\cite{McGovern:1992ix} and extensions of 
SM~\cite{Gunion:1990iv}-\cite{Bigi:1990kz}. 

Notice that our results   
for the coefficient $z_{\theta}$ is dominated by the pion-loop diagram.
The kaon-cloud contribution is smaller that the pion contribution
by a factor $\sim 1/28$: \eq z_\theta &=& z_\theta^\pi +
z_\theta^K \, = \, 1.42 \,, \\ z_\theta^\pi &=& 1.37\,, \quad\quad
z_\theta^K \, = \, 0.05 \,. \nonumber
\en
Using Eqs.~(\ref{h-theta}), (\ref{NDA1}), (\ref{NDA2}) in
Eq.~(\ref{tree+loop_D}), we can write for the electron-neutron
Schiff moment the following expression: \eq\label{Schiff_BSM}
S^{\prime} = \sum_{q=u,d,s}(s_q^E \, d_q^E \, + \, s_q^C \, d_q^C)
\, + \, s_W \, \Lambda_\chi \, C_W +  (\bar\theta\ s_{\theta}
\times 10^{-16}\ e \cdot cm)\, .
\en
where numerical values of the $s$-coefficients are \eq
\hspace*{-2cm} s_u^E &=& - 4.86 \ {\rm GeV}^{-2}\,, \hspace*{.5cm}
s_d^E \, = \, 17.85 \ {\rm GeV}^{-2}\,, \hspace*{.5cm} s_s^E \, =
\,  -0.09 \ {\rm GeV}^{-2}\,,  \nonumber\\[2mm] \hspace*{-2cm}
s_u^C &=& 0.41 \ {\rm GeV}^{-2}\,, \hspace*{.85cm} s_d^C \, =  \,
1.62 \ {\rm GeV}^{-2}\,, \hspace*{.7cm} s_s^C \, = \,  0.004 \
{\rm GeV}^{-2}\,,  \\[2mm] \hspace*{-2cm} s_W &=&  2.04 \ {\rm
GeV}^{-2}\,, \ \ \ \ \ \ s_{\theta}\ \, = 7.72 \,, \nonumber
\en
and where the separate contributions of pion and kaon meson-loop
diagram in Fig.~1a to the coupling $s_{\theta}$ are
$s_{\theta}^\pi = 7.62$ and $s_{\theta}^K = 0.1$.

Now, using Eq. (\ref{d-w}), we can extract upper limits for the
parameters $\bar\theta$, $d_q$, $\tilde h_q$, $d_q^E$, $d_q^C$ and
$C_W$ from the current experimental bound on neutron EDM in
Eq.~(\ref{EDM-exp-1}). These limits are shown in the Table~1. In
their derivation we assumed the dominance of each term at a time,
thus assuming the absence of significant cancellations between the
different terms contributing to the neutron EDM.

We note that, in comparison with the conventional valence quark
model which predicts for the neutron EDM
\begin{eqnarray}\label{VQM}
d_n = \frac{4}{3}d_d - \frac{1}{3}d_u \,,
\end{eqnarray}
the PCQM predicts a small but non-negligible contribution from the
strange quark EDM, which appears due to the K-meson loop diagram
in Figs.~1a and 1c. In Table~2 we show the predictions for the
neutron EDM of some other approaches. With the values of the
corresponding $z$-coefficients it is straightforward to derive
upper limits on the CP-violating parameters of the
Lagrangian~(\ref{eff56}) within the cited approaches. A general
summary is that our limits are less stringent. Notice that all the
approaches cited in Table~2, including the PCQM, were successful
in describing various properties of baryons, and yet they disagree
in the CP-violating sector, as they predict different  values of
the neutron EDM. This a manifestation of a strong hadronic model
dependence in the estimates of the neutron EDM. In view of that,
the existing EDM limits on fundamental CP violating parameters
could only be treated as a hint for model building rather than as
stringent limits.


\section{Summary and Conclusions}
\label{summary}

In this work we applied the perturbative chiral quark model to the
calculation of the electric dipole form factor and electric dipole
moment of the neutron. We parameterized a generic effect of CP
violation in terms of the effective quark electric dipole moments
and strong CP-violating quark couplings to the chiral fields,
which could have different sources at some more fundamental level.
We considered the relations between these effective parameters of
the quark-meson Lagrangian with more fundamental parameters of the
CP-odd quark-gluon Lagrangian. The latter includes the SM
$\theta$-term and the operators which originate from physics
beyond the SM, such as the chromoelectric and electric quark
dipole moment operators, as well as the Weinberg term. From the
existing limits on the neutron EDM we extracted the limits on the
individual effective quark EDMs and on the aforementioned more
fundamental CP violating parameters, adopting the approach based
on the so called na\"\i ve dimensional analysis. A general
conclusion we made from the comparison of these results with other
approaches is that our limits are significantly less stringent.
This could be treated as a manifestation of a strong dependence of
the predictions for the neutron EDM on the details of the hadronic
structure models.

\section*{Acknowledgments}

This work was supported in part by the DAAD under contract
415-ALECHILE/ALE-02/21672, by the FONDECYT projects 1030244,
1030254, 1030355, by the DFG under contracts FA67/25-3 and GRK683.
This research is also part of the EU Integrated Infrastructure
Initiative Hadronphysics project under contract number
RII3-CT-2004-506078 and President grant of Russia "Scientific
Schools"  No. 5103.2006.2. K.P. thanks the Development and Promotion
of Science and Technology Talent Project (DPST), Thailand for
financial support. C.D. thanks the Institute of Theoretical
Physics at the University of T\" ubingen, Germany, for its kind
hospitality. V.E.L. thank the Institute of the Departamento de
F\'\i sica at the Universidad T\'ecnica Federico Santa Mar\'\i a
of Valpara\'\i so, Chile, respectively, for their kind hospitality.

\newpage
\begin{table}\label{Table-1}
\caption{Upper limits on the CP-violating parameters of the Lagrangians
in Eqs.~(\ref{EQS1}), (\ref{eff56})
and CPVCs $\tilde h_q$; the high-energy constants $d_q^E$, $d_q^C$ and
from the current experimental bound on the neutron EDM
in Eq.~(\ref{EDM-exp-1}).
All numbers are given in units of $10^{-25}$ $e \cdot$cm.}

\vspace*{.25cm}
\def\arraystretch{1.}
\begin{center}
\begin{tabular}{|c|c|c|c|}
\hline
Parameter
& $u$ \hspace*{2cm}
& $d$ \hspace*{2cm}
& $s$ \hspace*{2cm}\\
\hline
$|d_q|$
& $2.0$  \hspace*{2cm}
& $0.5$  \hspace*{2cm}
& $105.0$ \hspace*{2cm} \\
$|\tilde h_q|$
& $0.4$  \hspace*{2cm}
& $0.4$  \hspace*{2cm}
& $12.6$ \hspace*{2cm} \\
$|d_q^E|$
& $1.3$  \hspace*{2cm}
& $0.4$  \hspace*{2cm}
& $68.6$ \hspace*{2cm} \\
$|d_q^C|$
& $6.4$ \hspace*{2cm}
& $2.9$  \hspace*{2cm}
& $174.7$ \hspace*{2cm} \\
\hline
$|\bar\theta|$ &
\multicolumn{3}{c|}{0.4\hspace*{2.4cm}} \\
\hline
$|C_W \, \Lambda_\chi|$ &
\multicolumn{3}{c|}{2.0\hspace*{2.4cm}} \\
\hline
\end{tabular}
\end{center}
\end{table}

\begin{table}
\caption{Theoretical estimates of $z$-coefficients
in Eq.~(\ref{d-w}) in various approaches.}

\vspace*{.25cm}
\def\arraystretch{1.}
\begin{center}
\begin{tabular}{|c|c|c|}
\hline
Parameter & Approaches & This work \\
\hline
$z_u^E$ & - (0.35 $\pm$ 0.15)~\cite{Olive:2005ru}\,;
                        -0.78~\cite{Abel:2004te}\,;
                         -1/3~\cite{He:1989xj}  & -0.48  \\
$z_d^E$ &     (1.4 $\pm$ 0.6)~\cite{Olive:2005ru}\,;
                         1.14~\cite{Abel:2004te}\,;
4/3~\cite{He:1989xj} & 1.77 \\ \hline
$z_s^E$ & -0.35~\cite{Abel:2004te} & -0.01\\ \hline
$z_u^C$ & 0.48~\cite{Hisano:2004tf}\,;
         (0.165 $\pm$ 0.075)~\cite{Olive:2005ru} \,;
          0.05~\cite{Gunion:1990iv}\,; & 0.10 \\
& 4/9 e $\simeq$ 0.135~\cite{He:1989xj} & \\
\hline
$z_d^C$ & 0.39~\cite{Hisano:2004tf}\,;
(0.33 $\pm$ 0.15)~\cite{Olive:2005ru} \,;
0.05~\cite{Gunion:1990iv}\,; & 0.21 \\
& 8/9 e $\simeq$ 0.270~\cite{He:1989xj} & \\ \hline
$z_s^C$ & 0.08~\cite{Hisano:2004tf}\, ;
          0.05~\cite{Gunion:1990iv}\,;
          0.03~\cite{Khriplovich:1996gj}\,;
          0.008~\cite{He:1990bu}\,
          &0.004 \\ \hline
$z^W$
          &  $\sim$ 0.085~\cite{Weinberg:1989dx} \, ;
             $\sim$ 0.017~\cite{Olive:2005ru}    \, ;
             $\sim$ 0.01~\cite{Bigi:1990kz}
          & 0.094\\ \hline
$z^{\theta}$ &
$2.7$~\cite{Baluni:1978rf}; $3.6$~\cite{Crewther:1979pi};
      $3.3$~\cite{Pich:1991fq}; $6.7$~\cite{Borasoy:2000pq};
      $3.0$~\cite{Musakhanov:1984qy};
          & 1.42\\
& $1.4$~\cite{Morgan:1986yy};
       $1.17$~\cite{McGovern:1992ix}; $2.4$~\cite{Pospelov:1999mv}&\\
\hline
\end{tabular}
\end{center}
\end{table}

\begin{figure}
\begin{center}
\epsfig{file=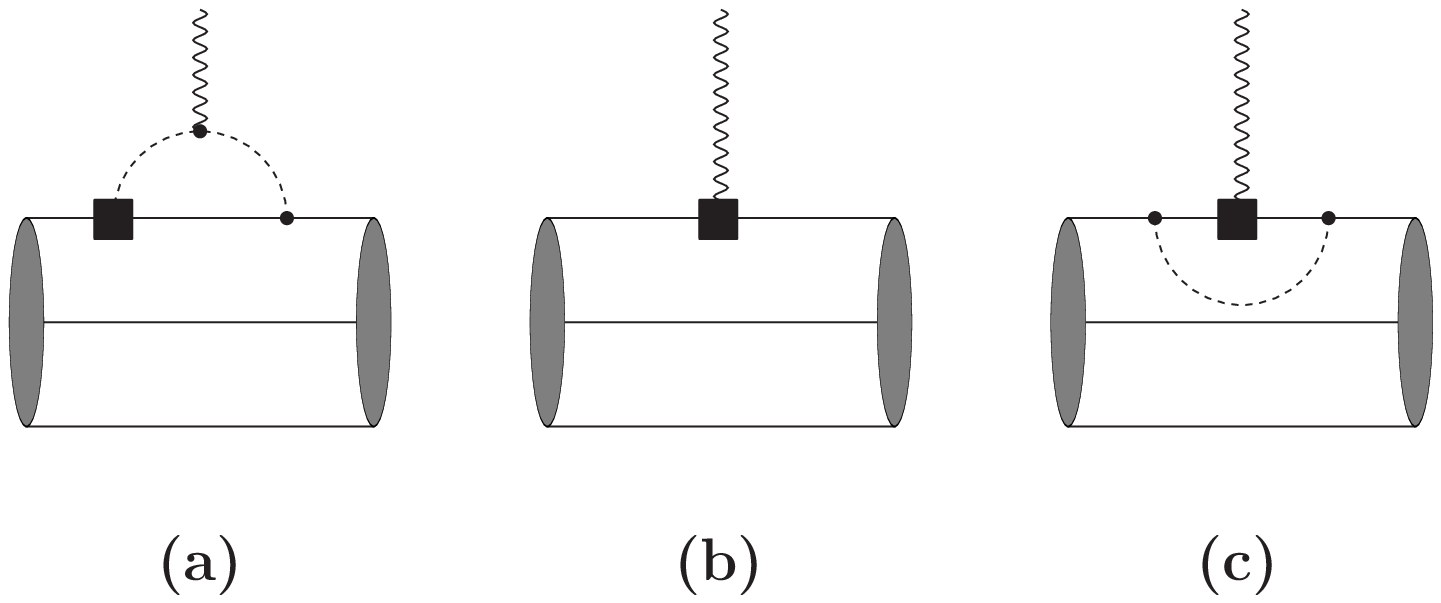, scale=0.8}
\end{center}

\vspace*{-4.5cm} \caption{The contributions to the neutron EDM:
``$h$-term" diagram (a), ``tree-level $d$-term" diagram (b) and
``one-loop $d$-term" diagram (c). The black filled squares
correspond to the CP-violating vertices of the
Lagrangian~(\ref{EQS1}) or the current~(\ref{CPcurrent}).}
\end{figure}

\newpage

\begin{figure}
\vspace*{1cm}
\begin{center}
\epsfig{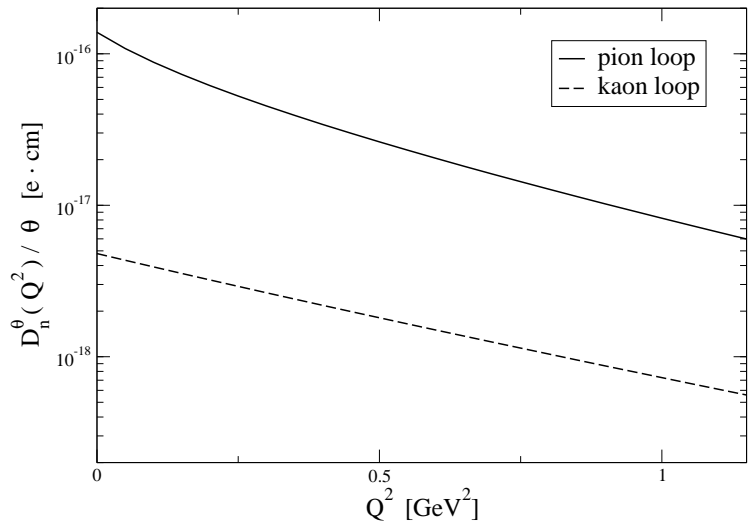}
\end{center}
\caption{The contributions from pion and kaon loops to the neutron
EDM form factor induced by strong CP-violating $\theta$-term,
where $D^{[\theta]}_n(Q^2) \equiv D^{[h]}_n(Q^2)$ at the limit
${\bf h_q} \equiv \theta \,\, \bar m \,\, {\bf I_{3 \times
3}}$\,.}
\end{figure}

\begin{figure}
\begin{center}
\epsfig{file=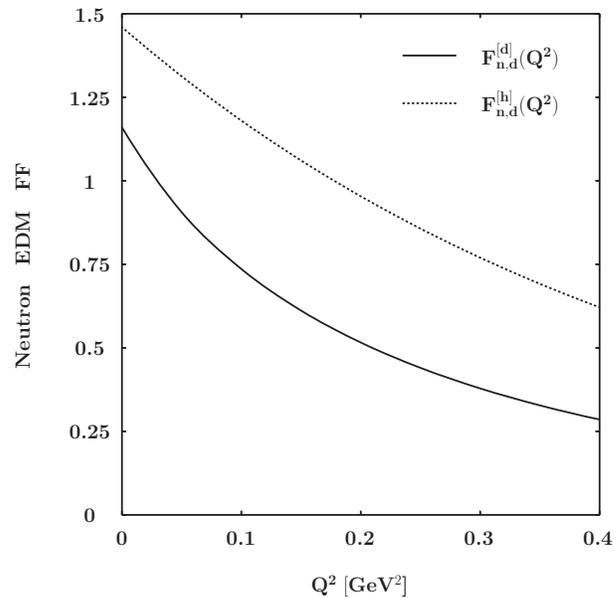, scale=0.45}
\end{center}
\vspace*{-4cm} \caption{The $d$-quark contribution to the neutron
EDM form factor via its intrinsic EDM (solid line) and CP-odd
coupling to mesons (dashed line). The plotted form factors are:
$F_{n,d}^{[d]}(Q^2)\, = \, \Delta_d^{[d]} \, F^{[d]}(Q^2)$ (solid
line) and $F_{n,d}^{[h]}(Q^2)\, = \, \Delta_d^{[h]} \,
F_d^{[h]}(Q^2)$ (dashed line).}
\end{figure}

\end{document}